**Kinetics of HEX-BCC Transition in a Triblock Copolymer in a Selective Solvent: Time Resolved Small Angle X-ray Scattering Measurements and Model Calculations.**

Minghai Li, Yongsheng Liu, Huifen Nie, Rama Bansil[*]

*Department of Physics, Boston University, Boston, MA 02215, USA*

Milos Steinhart

*Institute of Macromolecular Chemistry, Academy of Sciences of the Czech Republic, Heyrovsky Sq. 2, 162 06 Prague 6, Czech Republic*

**Abstract**

Time-resolved small angle x-ray scattering (SAXS) was used to examine the kinetics of the transition from HEX cylinders to BCC spheres at various temperatures in poly(styrene-*b*-ethylene-*co*-butylene-*b*-styrene) (SEBS) in mineral oil, a selective solvent for the middle EB block. Temperature-ramp SAXS and rheology measurements show the HEX to BCC order-order transition (OOT) at ~127 $^o$C and order-disorder transition (ODT) at ~180 $^o$C. We also observed the metastability limit of HEX in BCC with a spinodal temperature, $T_s$ ~ 150 $^o$C. The OOT exhibits 3 stages and occurs via a nucleation and growth mechanism when the final temperature $T_f < T_s$. Spinodal decomposition in a continuous ordering system was seen when $T_s < T_f < T_{ODT}$. We observed that HEX cylinders transform to disordered spheres via a transient BCC state. We develop a geometrical model of coupled anisotropic fluctuations and calculate the scattering which shows very good agreement with the SAXS data. The splitting of the primary peak into two peaks when the cylinder spacing and modulation wavelength are incommensurate predicted by the model is confirmed by analysis of the SAXS data.

[*] Author to whom correspondence should be addressed. Email: rb@bu.edu

**Introduction**

The transition of cylindrical micelles packed in a 2-dimensional hexagonal lattice (HEX) to spherical micelles on a BCC lattice has been studied with block copolymers quite extensively. In particular, the HEX cylinder to BCC sphere order-order transition (OOT) has been studied in di- and tri-block copolymer melts using small angle X-ray scattering (SAXS), small angle neutron scattering (SANS), transmission electron microscopy (TEM), rheology, and birefringence measurements [1, 2, 3, 4, 5, 6, 7, 8, 9, 10, 11]. The HEX cylinder phase is observed at lower temperature and BCC sphere phase emerges at a higher temperature. This transition is found to be thermally reversible. The cylinder to cubic spheres OOT has also been studied in block copolymer solutions [12, 13, 14, 15, 16, 17, 18]. The presence of a selective solvent can influence the temperature dependence of the phase diagram; for example in poly (styrene-*b*-isoprene) (SI) in diethylphthalate (DEP), a styrene-selective solvent, the cylinder to sphere OOT occurs upon cooling [15], whereas in the melt it appears on heating. SANS and SAXS measurements on sheared oriented samples of poly(ethylene propylene-*b*-ethyl ethylene) (PEP-PEE) diblocks [1], and on single grains of SI diblock melt [4] as well as on sheared oriented samples of SIS triblock [5, 6, 7, 8] reveal that the transition is epitaxial, with the axis of the cylinder becoming the <111> direction of the BCC lattice, and the (100) HEX planes becoming the (110) planes of BCC. However, to the best of our knowledge there are no measurements of the time-evolution of the HEX-BCC transition, although the kinetics of the reverse transition, BCC to HEX, which is much slower, has been investigated recently [19]. Although the phases are thermally reversible, the mechanism by which cylinders break up into spheres and form a BCC lattice is expected to be quite different than that involved in the reverse transition because the rate of formation of cylinders by merging spheres is much slower than the rate of breaking cylinders up.

It is generally accepted that the transition of HEX cylinders to BCC spheres involves the formation of undulated cylinders whose radii are modulated along the cylinder axis[1, 4, 5, 6, 7, 8, 16]. By allowing for anisotropic composition fluctuations Laradji et al [20] obtained modulated cylinders in their calculations of phase diagrams of a diblock copolymer melt showing the limits of metastability of the different ordered phases, and made predictions concerning the stability of gyroid and hexagonally perforated lamellar phases. Recently Ranjan and Morse[21] have re-examined the instability of the gyroid phase and the possibility of transitioning from BCC directly toward HEX instead of via a metastable perforated lamellar phase as suggested by Laradji et al.[20] These self consistent field calculations, as well as the time dependent Ginzburg Landau (TDGL) calculations of the kinetics of the HEX to BCC transition also support the occurrence of modulated cylinders[22]. Matsen[23] used self-consistent mean field theory to examine the pathway of the cylinder to sphere epitaxial transition and showed a nucleation and growth mechanism with strong fluctuation effects due to the small energy barrier of the transition. He also noted a narrow window in which spinodal mechanism would occur. Such rippled cylinders were seen by Ryu and Lodge[6, 7] using TEM and SAXS in an oriented SIS melt. Recently Bendejacq et al [24] have obtained high-resolution TEM images of rippled cylinders in poly(styrene-*b*-acrylic acid) (PS-PAA) diblocks dispersed in water, which enabled them to measure structural parameters, such as the wavelength of the ripple ($\lambda$), the radius of the core ($R_c$) and the height ($h$) of the brush (related to the amplitude of the fluctuation). From these measurements they concluded that the ratio of the height of the cylindrical PAA brush to its core radius determines the separation between undulating cylinders, straight cylinders and spheres. Specifically they found that straight cylinders are found in the case $h/R_c \leq 1.8$, undulating cylinders between $1.8 < h/R_c < 2.0$ and spheres above $h/R_c \geq 2.0$. Their measurements clearly support the criterion of a critical curvature as driving the transition from cylinders to spheres. A theoretical study on the same system by Grason et al[25] shows that the modulated cylinder is a metastable state in the cylinder to sphere transition under a certain range of charge and salt concentration where the sphere state is the thermodynamically favored stable state.

The breakup of a cylinder into spheres without any underlying lattice has been studied extensively in the so-called "pearling instability", according to which the amplitude of a transverse wave along the length of the cylinder grows causing the cylinder to break up into droplets ("pearls"). This is observed in the classical experiments of Rayleigh[26, 27] where a column of liquid pinches into drops, as well as in lipid vesicles in an optical laser tweezer trap[28, 29]. The growth of the instability involves the competition between the surface energy and bending elasticity of the cylinder[27, 28, 30, 31, 32].

In the block copolymer case the transition involves both the breaking of cylinders to spheres and the epitaxial transformation of the underlying lattice. While previous experimental and theoretical studies provide insight into the epitaxial mechanism, and rippled cylinders have been observed by TEM, there are no measurements of the time evolution of the transformation, nor are there any reports of a formalism to calculate the azimuthally averaged scattering intensity from rippled cylinders in an un-oriented system. Time-resolved SAXS provides a convenient probe of the transformation kinetics. Unlike the direct visualization afforded by TEM and atomic force microscopy (AFM) methods, extraction of spatial structural information from SAXS data is not so direct. One needs a geometrical model so as to be able to correlate features in the momentum-space scattering data with the real-space morphology of the system.

In Part I of this paper we report synchrotron based time-resolved SAXS measurements of the kinetics of the transformation from HEX to BCC in the triblock copolymer of poly(styrene-*b*-ethylene-*co*-butylene-*b*-styrene) (SEBS) in mineral oil, a selective solvent for the middle PEB block. This system forms a network of micelles with PS in the cores and the solvated PEB chains forming loops and bridges. Because the solvent is poor for the minority PS block it further enhances the microphase separation tendency due to the incompatibility of PS and PEB blocks. At a concentration of 45% the system exhibits a HEX phase at lower temperatures than the BCC phase. This behavior is similar to that of SEBS in the melt. We examine the kinetics of HEX to BCC transition for different values of the final temperature, as well as the HEX to disorder spherical micelle transition. From an analysis of these data we obtain detailed insight into the mechanism of the transition and the temperature dependence of the kinetics. In the second part of this paper, we develop the structural model of rippled cylinders to calculate the scattering and compare with the experimental results.

**Experimental Section**

**Materials.** A 45% w/v solution of SEBS triblock copolymer (Shell Chemicals, Kraton G1650) with a molecular weight $M_n$ of 100,000 Daltons, polydispersity $M_w/M_n$ of 1.05, styrene fraction 28 wt%, and E:B ratio 1:1 was prepared in mineral oil (J.T. Baker) which is selective to the middle PEB block. Methylene chloride was used as a co-solvent to make a homogeneous solution and then was removed by evaporation until no further change in weight was observed.

**Rheology.** The dynamic storage and loss moduli $G'$ and $G''$ were measured as a function of temperature on an AR-G2 rheometer (TA instruments) at the Hatsopoulos Microfluids Laboratory at MIT. We used an angular frequency $\omega$ of 1 rad/s and strain $\gamma_0$ of 2% as these parameters correspond to the linear viscoelastic regime. For the heating process a controlled temperature ramp rate of 1 °C/min was used.

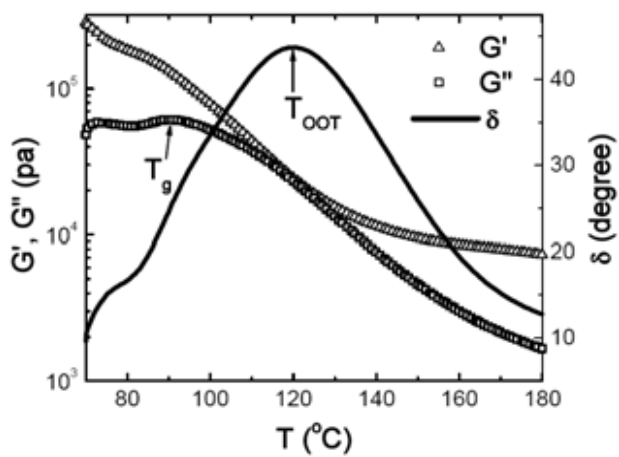

Figure 1. Temperature dependence of the dynamic shear moduli $G'$ and $G''$ (at $\omega$ = 1 rad/s, and strain $\gamma_0$ = 2%) from SEBS 45% in mineral oil at a heating rate of 1 °C/min.

The temperature dependence of $G'$, $G''$ and $\delta = \tan^{-1}(G''/G')$ upon heating is shown in Figure 1, and reveal a glass transition at ~ 90 °C and an order-order transition $T_{OOT}$ at ~120 °C. The data agree with low frequency measurements reported on a mixture of S-EB and SEBS[11], and show similar behavior to that of the randomly oriented sample of SIS in reference 7. Because the sample is randomly oriented, $G'$ and $G''$ of the cylindrical phase are higher than that of the spherical phase. Hysteresis effects are observed on cooling (data not shown), as is usually expected with these materials.

**Atomic Force Microscopy.** A Model 3100 AFM (Digital Instruments, Santa Barbara, CA) attached to a Nanoscope IIIa controller with an electronic extender box at Boston University Photonics Center was used for the present studies. The sample of SEBS 45% in mineral oil was spin cast on a silicon wafer by diluting in toluene which evaporates during the spin casting process. The spin cast sample was annealed at 110 °C for 24 hours. Just before AFM imaging the sample was quick-frozen in liquid $N_2$ to preserve the high temperature morphology and imaged at ambient temperature using tapping mode. The height image from the AFM measurement shown in Figure 2 reveals a well-ordered HEX morphology with a d-spacing of 35 nm between neighboring cylinders. The radius of the cylinder is estimated to be 10 nm. Note, that due to tip broadening this is an overestimate of the actual cylinder radius.

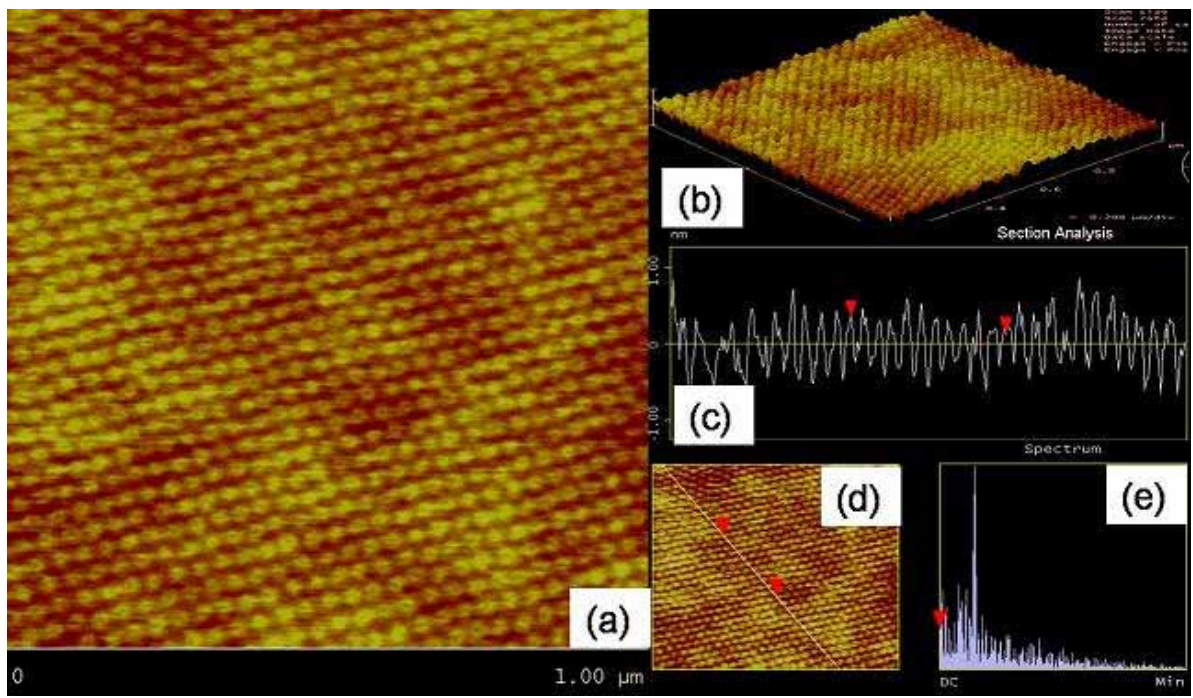

Figure 2(a). Room temperature AFM height image (1μm x 1μm) from SEBS 45% in mineral oil annealed at 110 °C and quick frozen to preserve the annealed morphology shows hexagonally packed cylinders. The 3-dimensional rendering in (b) shows that the cylinders are oriented perpendicular to the substrate. The contrast in the image is due to the differential hardness of the glassy PS cores (bright) and the soft PEB matrix (dark). A section analysis gives the height profile (c) along the line shown in the lower image (d) indicating that the spacing between the cylinders is 35 nm. The power spectrum of the image (e) exhibits good order along the line indicated in the image shown in (d).

**Small Angle X-ray Scattering.** Time-resolved SAXS experiments were conducted at beamline X27C of the National Synchrotron Light Source (NSLS) of Brookhaven National Laboratory, using X-ray of wavelength $\lambda$ = 0.1366 nm (9.01 keV) with energy resolution $dE/E$ =

1.1%. The scattering intensity was recorded on a 2 dimensional MAR CCD detector with an array of 1024 x 1024. For the solution samples used here, the scattering patterns were isotropic, so an azimuthal average was done to obtain the scattered intensity $I(q)$ as a function of the scattering wavenumber, $q$ over the range $0.1 < q < 3$ nm$^{-1}$. The sample was loaded into a custom designed cell made of a copper plate with a 0.6 cm-diameter hole covered with two thin flat Kapton windows. A custom-designed computer controlled Peltier heater/cooler module connected to the sample cell was used to change the temperature either rapidly (temperature jump) or at a constant rate (temperature ramp). The desired temperature was reached within 1 minute with the Peltier module. Typically, the scattering intensity profile $I(q)$ was recorded for approximately 10 s per frame (includes data acquisition time and the time to read the array), and the total time of each run was 1 to 2 hours. All scattering data were corrected by normalizing by the incident beam intensity, and subtracting the scattering from the solvent. This procedure allows us to compare the relative intensity from different frames following a temperature jump or ramp, but it is important to note that the intensity data are not calibrated against a standard and hence do not give the absolute intensity. It is also important to note that the cell used here allows has flexible Kapton windows. Details of the experimental set up and data processing are described in our previous work on disorder to BCC kinetics in SEBS triblock copolymer solution in mineral oil[33].

**Results and Discussion**

**Part I: Small Angle X-ray Scattering Experiments**

Figure 3 shows long-time averaged SAXS data from the 45% SEBS in mineral oil sample at 110 $^{\circ}$C and 155 $^{\circ}$C averaged for 20 min. These data clearly confirm that at 110 $^{\circ}$C the sample is in the HEX phase (peaks at relative positions of $1:\sqrt{3}:\sqrt{4}:\sqrt{7}$), while at 155 $^{\circ}$C it is in the BCC phase (peaks at relative positions of $1:\sqrt{2}:\sqrt{3}:\sqrt{4}:\sqrt{5}:\sqrt{6}:\sqrt{7}$).

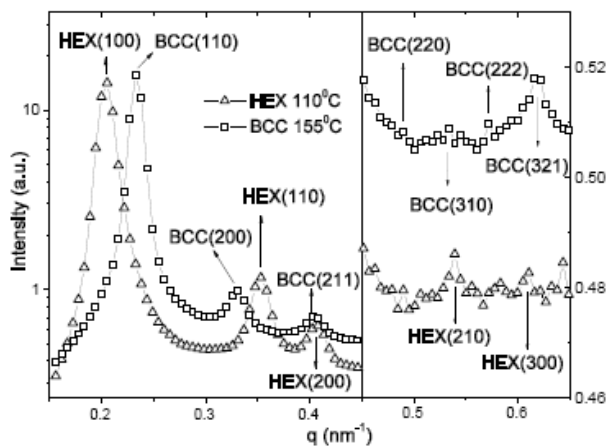

Figure 3. Long-time averaged SAXS data (in arbitrary units (a.u.)) from 45% SEBS in mineral oil at 110 $^{\circ}$C and 155 $^{\circ}$C. The first 5 peaks for HEX at 110 $^{\circ}$C and the first 7 peaks for BCC at 155 $^{\circ}$C are marked. The plot is divided into two parts with an enlarged intensity scale for the high-$q$ region to clearly display the higher peaks.

**Identification of HEX to BCC Transition by Temperature Ramp Measurement.** To determine the OOT temperature, SAXS data were acquired while the sample was being heated at a constant rate of 1 $^{\circ}$C/min from 70 $^{\circ}$C to 180 $^{\circ}$C, as shown in Figure 4.

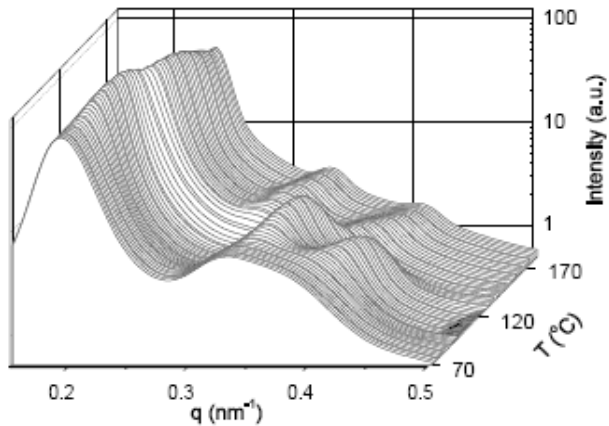

Figure 4. Time evolution of the scattered intensity $I(q)$ during heating of the SEBS sample from 70 °C to 180 °C at a constant rate of 1 °C/min.

Initially at 70 °C the sample is in a glassy state and a HEX phase can be clearly identified around 90 °C. The relative positions of the peaks are characteristic of the HEX structure in the temperature range of 90 - 120 °C and of the BCC structure at higher temperatures, with a clear transition in the vicinity of 120 - 130 °C. The appearance of a peak ($\sqrt{2}$ peak) at $q_2 = \sqrt{2}\, q_1$, where $q_1$ denotes the primary peak position, is indicative of the transition from the HEX to BCC phase. To identify the transition temperature we plot the peak intensity ($I_1$ and $I_2$) and position ($q_1$ and $q_2$) of the first two Bragg peaks as a function of the temperature, as shown in Figure 5. The peak intensity, position and width of the primary peak were determined by a fitting procedure described in previous publications from our group[34]. The values obtained by the fitting procedure are in excellent agreement with parameters determined by direct inspection of the data. The peak position is determined to an accuracy of 0.004 nm$^{-1}$. The parameters for the second peak were directly obtained from the data, because the peak is weak and that makes the fitting method unreliable.

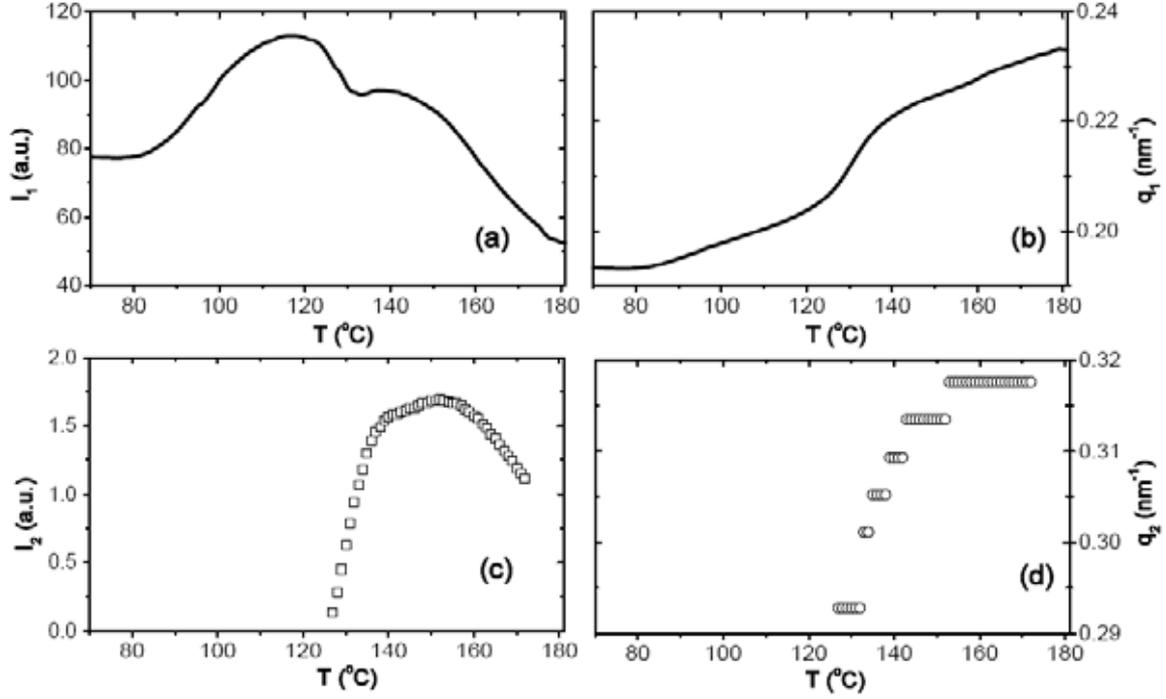

Figure 5. Temperature dependence of the first two Bragg peaks measured from the temperature ramp data shown in Fig. 3. (a) Primary peak intensity $I_1$, (b) position $q_1$. (c, d) The same parameters for the √2 peak. The discrete jumps in the peak position are due to the fact that one pixel at the detector corresponds to 0.004 nm$^{-1}$.

We observe that the intensity of the primary peak reaches a local minimum at ~127 °C, and the √2 peak (characteristic of the BCC structure) first appears at ~127 °C. From this we identify that the HEX→BCC order-order transition temperature $T_{OOT}$ ~ 127 ± 2 °C. Above this temperature $I_2$ increases rapidly. We also found that the intensity, $I_3$, of the √3 peak (not shown) decreases above this temperature. These changes indicate the conversion of the HEX state to the BCC state. We observe that $q_1$ increases with temperature, implying that the lattice constant decreases with increasing temperature. Similar shift in peak position with increasing temperature, has also been noted in earlier experimental work [3, 4, 7, 9, 19, 35] and also been predicted by theory[36].

The intensity $I_2$ reaches a maximum at 150 °C, and as discussed later, we identify this temperature as the spinodal temperature ($T_s$) corresponding to the metastability limit of HEX in BCC. We also observed that at about 180 °C the intensity $I_2$ decreases rapidly, and the peaks become broader, indicating the onset of an order-disorder transition. It is important to know that the temperature ramp method over-estimates the transition temperatures since the results depend on the rate of heating.

**Kinetics of the HEX→BCC Transition.** To study the kinetics of this order-order transition we made time-resolved SAXS measurements following a temperature jump (T-jump) from a sample annealed at a fixed initial temperature $T_i$ = 110 °C in the HEX phase to various final temperatures $T_f$ above $T_{OOT}$. The kinetics was measured for about 2 hours. It took about 60 - 160s for the system to respond to the T-jump and reach the desired final temperature. The temperature equilibration time, $t_0$ depends on the magnitude of the temperature jump, defined as $\Delta T = T_f - T_i$, as shown later in Table 1. Typical SAXS results for the early stages of the T-jump with $T_f$ = 135 °C are shown in Figure 6.

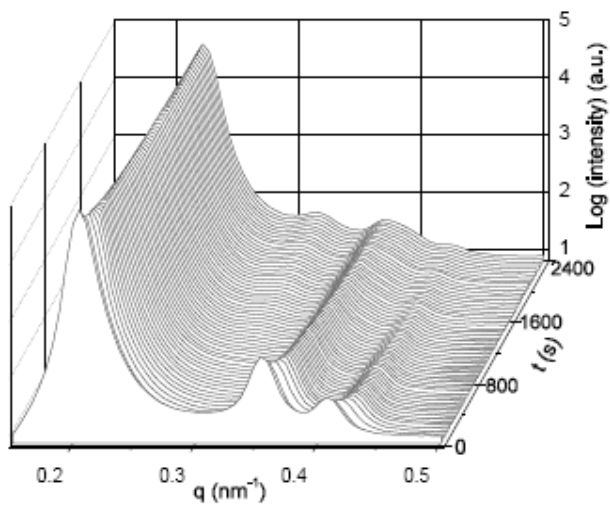

Figure 6. Time evolution of the SAXS intensity following a T-jump from 110 °C to 135 °C.

**Analysis of SAXS Data.** To follow the details of the time evolution we analyzed the primary peak's position ($q$), intensity ($I$), and width ($w$) as a function of time. Figure 7 shows the time evolution of these parameters for the T-jump from 110 $^\circ$C to 135 $^\circ$C. Immediately following the T-jump, the temperature of the sample changes rapidly and reaches the final temperature, $T_f$ in time $t_0 = 60$ s (see Figure 7c). During this initial time the structure is that of HEX lattice and the intensity drops very rapidly, in response to the rapid change in temperature. Further isothermal time evolution exhibits three stages:

Stage I ($t_0 < t < t_1$), where the structure still shows peaks characteristic of the HEX lattice, but the peak positions shift rapidly to higher values, indicating that the cylinders are moving closer together. During this stage the primary peak intensity $I_1$ (of HEX (100) peak) decreases, and the slope of $I_1$ versus $t$ graph shows a sharp change at $t_1 = 120$ s;

Stage II ($t_1 < t < t_2$), in this period the primary peak intensity first grows then decreases, reaching a second local minimum at $t_2 = 1400$ s;

Stage III ($t > t_2$), during which the intensity of the primary peak increases monotonically, reaching a stable value for the BCC phase eventually.

The separation into three stages is also supported by the non-monotonic behavior of the peak width, $w_1$ which has a minimum at $t_1$ and maximum at $t_2$. The peak position $q_1$ increases rapidly up to $t_1$ and slowly thereafter, becoming more or less stable after $t_2$. A similar approach was used by Sota et al[19] in the analysis of the transition of BCC to HEX for a shallow quench. They also observed three stages after the temperature incubation time, with two steps corresponding to the phase transition period and the last one to the growth of the HEX structure.

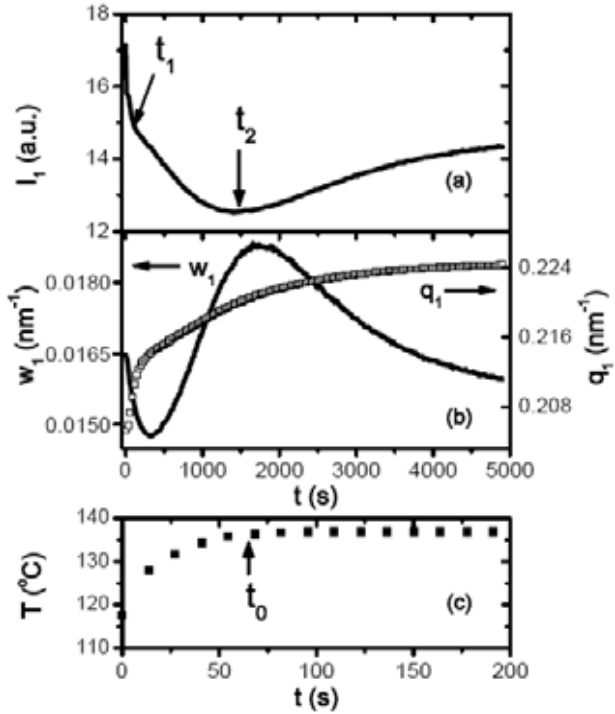

Figure 7. Time evolution of (a) the intensity ($I_1$), (b) the position ($q_1$) and width ($w_1$) of the primary Bragg peak following a T-jump from 110 °C to 135 °C (see the SAXS data shown in Figure 6). (c) The temperature equilibration during the very early stages of the jump. Note the temperature data is only shown for the first 200 seconds, although the temperature was recorded throughout the experiment. The times $t_0$, $t_1$ and $t_2$ (see text for definition) indicating the different stages in the time evolution are labeled.

Further support for the onset of BCC structure is obtained from analyzing the $\sqrt{2}$ peak corresponding to the (200) reflection of the BCC lattice which appears around $t_1$ and is clearly identifiable around $t_2$ (as shown in Figure 8a). The intensity $I_2$ increases after $t_1$, while its width narrows as shown in Figure 8b.

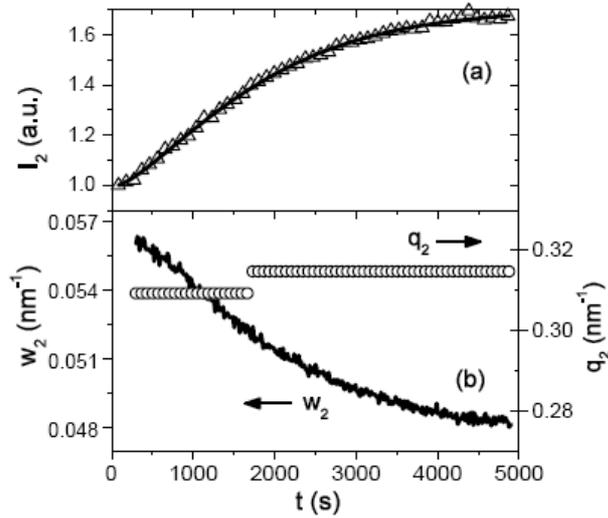

Figure 8. The time evolution of (a) the intensity ($I_2$), and (b) the width ($w_2$) and position ($q_2$) of the $\sqrt{2}$ peak following a T-jump from 110 °C to 135 °C. The solid line in (a) is the fit to Eq. (1) describing a stretched exponential growth of the BCC phase, as described later in the text. The position $q_2$ is determined to an accuracy of ~ 0.004 nm$^{-1}$.

The first stage corresponds to a pre-transitional incubation period where HEX cylinders move closer together; the second represents the phase transition from HEX cylinders to BCC spheres, i.e., the cylinders are modulated and modulation amplitude grows until cylinders break up into the spheres in BCC symmetry; and the third represents the growth of the BCC sphere phase. The presence of the incubation period indicates a nucleation and growth mechanism, instead of spinodal decomposition.

**Dependence of the Kinetics on the Depth of the T-jump.** Figure 9 shows the data for the peak parameters for the various T-jump measurements.

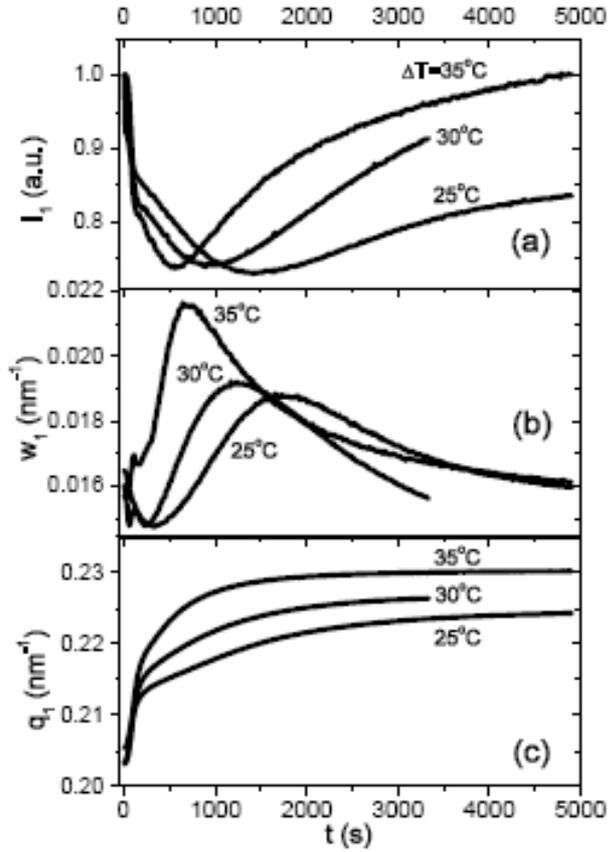

Figure 9. Time evolution of peak intensity ($I_1$), width ($w_1$), and position ($q_1$) of the primary peak of SAXS data for various T-jumps ($\Delta T$ = 25, 30 and 35 °C as indicated). The initial temperature was fixed at 110 °C for all T-jumps. All T-jumps show a 3-stage time evolution following the initial temperature equilibration period.

The T-jumps with $\Delta T$ = 30 and 35 °C show similar behavior as described above for $\Delta T$ = 25 °C with three time regimes. In Fig.9, the intensity is normalized by the initial intensity at $t = 0$. and the figure shows that the intensity at late time (around 5000 s) increases with increasing $\Delta T$, while the peak width narrows. The intensity depends on the amount of transformation material as well as on the size of growing microdomains. In the limited duration of our experiment the coarsening process is not complete. For the larger T-jump (with larger $\Delta T$), the coarsening process is faster and hence the domains are bigger in size, giving a higher intensity. The larger size of domain for larger T-jump is also confirmed by the narrower peak width (shown in Fig.9b). The peak position data (Fig.9c) is consistent with previous measurements[3, 4, 7, 9, 19, 35] which show that lattice constant decreases with increasing temperature in both HEX and BCC phase. The time $t_1$ before which the HEX cylinders move closer together is almost independent of temperature; the time $t_2$ decreases with increasing temperature jump (see Table 1).

Table 1. The temperature dependence of different time regimes and the position of the primary peak for various T- jumps in the HEX to BCC transition.

| $\Delta T$ (°C) | $t_o$ (s) | $t_1$ (s) | $t_2$ (s) | $q_1$(BCC)/$q_1$(HEX) | $q_1$ (BCC) (nm$^{-1}$) |
|---|---|---|---|---|---|
| 25 | 60 | 120 | 1400 | 1.09 | 0.224 |
| 30 | 60 | 150 | 930 | 1.11 | 0.226 |
| 35 | 100 | 130 | 560 | 1.13 | 0.230 |
| 45 | 120 | 200 | | 1.14 | 0.234 |
| 120 | 160 | t′ = 1500 s | | 1.20 | 0.255 |

This indicates that with increasing $\Delta T$ the transformation occurs faster, as is usually expected due to the increased thermodynamic driving force for larger jump in temperature. The temperature dependence of ($t_2$ - $t_1$), the transition period, for the $\Delta T$ = 25, 30 and 35 °C jumps is shown in Figure 10 and extrapolated to zero by linear fitting. The transition period vanishes at around $\Delta T = 40\ ^{\circ}$C corresponding to $T_f = 150\ ^{\circ}$C, which indicates that the transition mechanism is different above and below 150 $^{\circ}$C. We estimate that this temperature is in the vicinity of the limit of metastability of HEX in BCC, i.e. the spinodal temperature ($T_s$). At this temperature we also observed that $I_2$ is a maximum in the temperature ramp measurement (see Figure 5). To make a semi-quantitative comparison with theory we examine the ratio of $T_s / T_{OOT}$. Since we could not find any calculated phase diagram for a triblock copolymer in a selective solvent we make a qualitative comparison with predictions for diblocks. As pointed out by Matsen et al[37] the melt of an ABA triblock exhibits similar phase diagram as the AB/2 diblock formed by snipping the triblock in half, but differ in mechanical property due to the formation of bridges of the middle B block between two outer A domains. The selectivity of mineral oil to middle block PEB further enhances the microphase separation tendency between PS and PEB block, therefore our triblock SEBS solution has similar behavior as the melt of S-EB, although $\chi N$ for phase boundaries will be at slightly different values than the diblock prediction. Matsen[23] predicted a narrow window for diblock copolymer melt in which the mechanism would be that of spinodal decomposition. In his calculation, for a diblock copolymer melt of composition fraction $f = 0.28$, the OOT of HEX to BCC is at $(\chi N)_{OOT} \approx 16.4$ and the spinodal is at $(\chi N)_s \approx 15.3$. We find that the ratio of the $T_s/T_{OOT}$ obtained from the SAXS data for SEBS in mineral oil is 423K/400K $\approx$ 1.06 which agrees quite well with Matsen's prediction[23] of $(\chi N)_{OOT}/(\chi N)_s \approx 1.07$ for a diblock melt with $f = 0.28$ (assuming that $\chi \sim 1/T$, where $T$ is the absolute temperature).

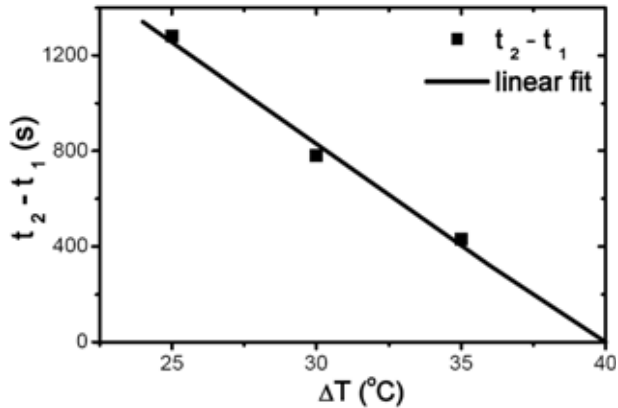

Figure 10. The dependence of $t_2 - t_1$, the transition period for HEX to BCC transformation on the depth of the T-jump ($\Delta T$). The straight line is a linear fit to the data which extrapolates to zero at $\Delta T = 40\ ^{\circ}$C corresponding to the limit of metastability for the HEX phase at 150 $^{\circ}$C.

**Growth of the BCC Structure Following a T-jump.** The time evolution of $I_2$ can be used as a measure of the growth of the BCC phase. The peak intensity $I_2$ increases with time approaching a steady value finally. The time evolution of $I_2$ could be fit by the stretched exponential formula:

$$I(t) - I(t_0) = (I(t_\infty) - I(t_0))(1 - e^{-((t-t_0)/\tau)^n}). \quad (1)$$

Figure 11 shows the normalized intensity of the √2 peak, defined as $[I(t) - I(t_0)] / [I(t_\infty) - I(t_0)]$, and the results of the fit to Eq. (1). The normalized data almost coincide for the jumps with $\Delta T = 25, 30\ ^\circ C$ but are identifiably different for $\Delta T = 35$ and $45\ ^\circ C$. The fitting parameters $\tau$ and $n$ are listed in Table 2. The characteristic time $\tau$ decreases with increasing $\Delta T$ as expected. For a T-jump above the spinodal line, $\Delta T = 45\ ^\circ C$, $\tau$ is much smaller, indicative of a faster process of the transition due to the larger driving force for deeper jumps. The exponent $n$ is close to 1 for $\Delta T = 35$ and $45\ ^\circ C$ (corresponding to exponential behavior), and departs slightly from exponential for the shallower jumps (1.2 -1.3).

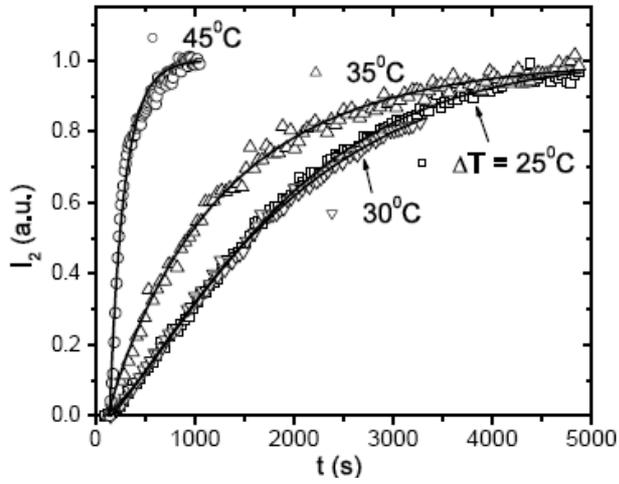

Figure 11. The time evolution of the normalized intensity of the √2 peak data for various T-jumps as indicated by the different symbols. The results of a stretched exponential fit (Eq. (1)) to the data are shown with the solid lines. The normalized data for $\Delta T$ = 25, 30 °C are very close together indicating a very weak temperature dependence in this temperature range.

Table 2. The parameters for the stretched exponential fit to Eq. (1) of √2 peak intensity $I_2$ for different T-jumps.

| $\Delta T$ (°C) | $\tau$ (s) | $n$ |
|---|---|---|
| 25 | 1918 | 1.3 |
| 30 | 1911 | 1.2 |
| 35 | 1185 | 0.9 |
| 45 | 147 | 0.9 |

The overall growth curves are similar to the predictions of the TDGL calculations[22] for the HEX to BCC transformation via the growth of anisotropic fluctuations. The interpretation of the SAXS data in terms of the growth of anisotropic fluctuations will be discussed in the second part of this work. The stretched exponential fit to the data for the shallow jumps which are below the metastability limit is consistent with a nucleation and growth mechanism, usually described by the Avrami equation[38] which has the same form as Eq. (1). It is important to note that in this transition ripples along the cylinder nucleate and form spherical micelles, as has been described in detail by Matsen[23].

**Kinetics of a T-jump Above the Spinodal.** For a deep jump with $\Delta T = 45\ ^oC$ (i.e. to a temperature above $T_s$), we observed a qualitatively different behavior than for the shallow jumps below the spinodal, as shown in Figure 12.

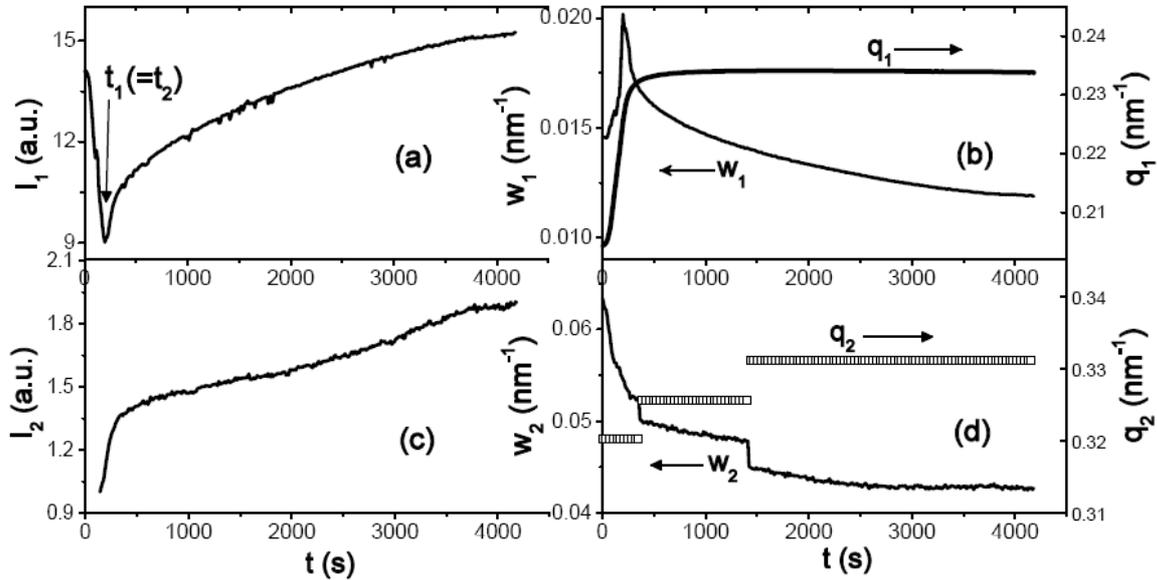

Figure 12. Time evolution of the intensity $I$, width $w$ and position $q$ of the primary peak and the $\sqrt{2}$ peak of SAXS data for T-jump from 110 $^oC$ to 155 $^oC$.

For the $\Delta T = 45\ ^\circ\text{C}$ jump the two times $t_1$ and $t_2$ cannot be distinguished and no intermediate stage could be detected. The transformation from the HEX cylinders to BCC spheres occurs via the mechanism called Model C in the Hohenberg-Halperin[39] classification scheme which involves both a non-conserved field (due to the symmetry changing transition) and a conserved field (composition of block copolymer as well as concentration are both conserved). Although the conservation condition is similar to the spinodal decomposition in a binary polymer blend[40], the symmetry breaking is unique to the block copolymer. The symmetry-breaking feature is similar to that observed in continuous ordering in metallic alloys, but there is no conservation condition in that case[41]. Unlike the Cahn-Hilliard equation[42] used for describing spinodal decomposition in a polymer blend, there is no simple analytical expression for predicting the time evolution of the scattering function for Model C. As expected the transition for the deeper jump with $\Delta T = 45\ ^\circ\text{C}$ occurs faster than for the shallower jumps with $\Delta T < 40\ ^\circ\text{C}$ due to the larger thermodynamic driving force.

**Jump from HEX to Disorder State Exhibits a Transient BCC Phase.** We also made a very deep T-jump measurement from the HEX phase at 110 $^\circ$C to 230 $^\circ$C at which the sample is eventually disordered. For this very deep jump with $\Delta T = 120\ ^\circ\text{C}$ as shown in Figure 13, the peak intensity $I_1$ decreases rapidly during the time $t_0$ in which the sample temperature equilibrates. After this time a BCC phase can be identified, which persists for about 1500 seconds. In this time interval the $\sqrt{2}$ peak appears and grows in intensity and gets slightly narrower, while the primary peak changes little in intensity or width.

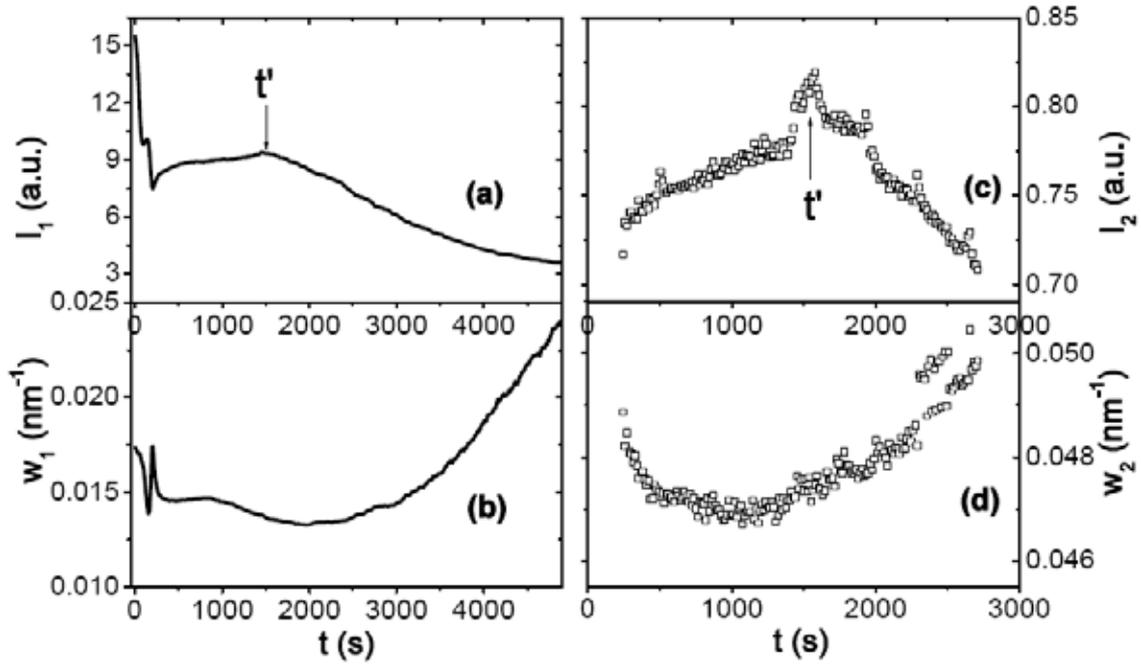

Figure 13. Time-evolution of the intensity and width of the primary peak and √2 peak for a very deep T-jump from 110 °C to 230 °C. The HEX structure transforms initially to a BCC state which persists for about 1500 seconds. The disordering time $t' = 1500$ s, after which the system transforms from BCC to a disordered micellar sphere state, is indicated.

After 1500 seconds the sample becomes disordered as evidenced by the decreasing intensity and broadening of both peaks. Hence, we conclude that the HEX cylinders first undergo an order-order transition forming a transient BCC sphere phase and then order-disorder transition occurs at around 1500 seconds identified as $t'$ in Figure 13. Interestingly, in the reverse transition from disordered spheres to HEX cylinders reported by Sota et al [43] a transient BCC sphere state was seen for a shallow quench below $T_{OOT}$. A transient BCC state was also observed in the disorder to FCC transition in a SI block copolymer in tetradecane, an isoprene selective solvent[34].

**Geometrical Characteristics of the HEX Cylinder and BCC Sphere Phases.** From Table 1 we note that the ratio $q_1$(BCC)/$q_1$(HEX) = $d_{100}$/ $d_{110}$ (where $d_{100}$ and $d_{110}$ denote the principal lattice spacing of the HEX and BCC phases = $2\pi/q_1$)[7] increases with increasing $\Delta T$. For large $\Delta T$, the ratio is bigger than the theoretical prediction of 1.08 for melts[36], and is also larger than the values reported from experiments in melts [4, 7]. From the positions of the primary Bragg peak we can estimate the principal lattice spacing $d^* = 2\pi/q_1$, and find that it varies from 30.9 nm for the HEX cylinder structure to 25.3 - 28.0 nm for the sphere BCC structure. The position of the first minimum ($q_{min}$) of the form factor can be related to the core radius via $r = 4.49/q_{min}$ for sphere and $r = 3.83/q_{min}$ for cylinder. We determined this minimum by examining the SAXS data on a greater magnification than shown in Figure 2. The radius of the cylindrical domain estimated using this relationship is about 8 nm (for $T$ = 110 °C), while the radius of sphere is about 10 nm (for $T$ = 135 °C). We note that the cylinder radius value obtained from the AFM image of the HEX phase is a little larger than that calculated from the SAXS data as expected due to tip broadening effects. The position $q_{min}$ in the SAXS data shifts continuously to lower value as time increases reflecting the increase in the radius from cylinder to sphere. The average end-to-end length of a PS chain of the SEBS triblock treated as Gaussian is $L_{PS} = a_{PS} N_{PS}^{0.5}$ = 8 nm using $a_{PS}$ = 0.71 nm and $N_{PS}$ = 134 as the number of PS monomers in each PS chain[44]. This length is very close to the radius of the cylinder but smaller than the radius of the sphere, suggesting that in the sphere phase there may be some solvent in the core and/or the chain may be stretched. This is plausible because solvent selectivity decreases with increasing temperature.

**Part II: Model Calculation of Scattering Intensity**

In this section we discuss a geometrical model to calculate the scattering from the rippled cylinders that form in the process of the HEX to BCC transition. This model is applicable to explaining the temperature jumps beyond the spinodal, where the cylinders are unstable with respect to modulation, and thus ripples form over the entire length of the cylinders which are correlated with their neighbors. Figure 14 schematically shows the geometrical model for the transition from HEX cylinders to BCC spheres (adapted from Laradji et al[20]) with seven unmodulated cylinders in 2-d HEX lattice as initial state.

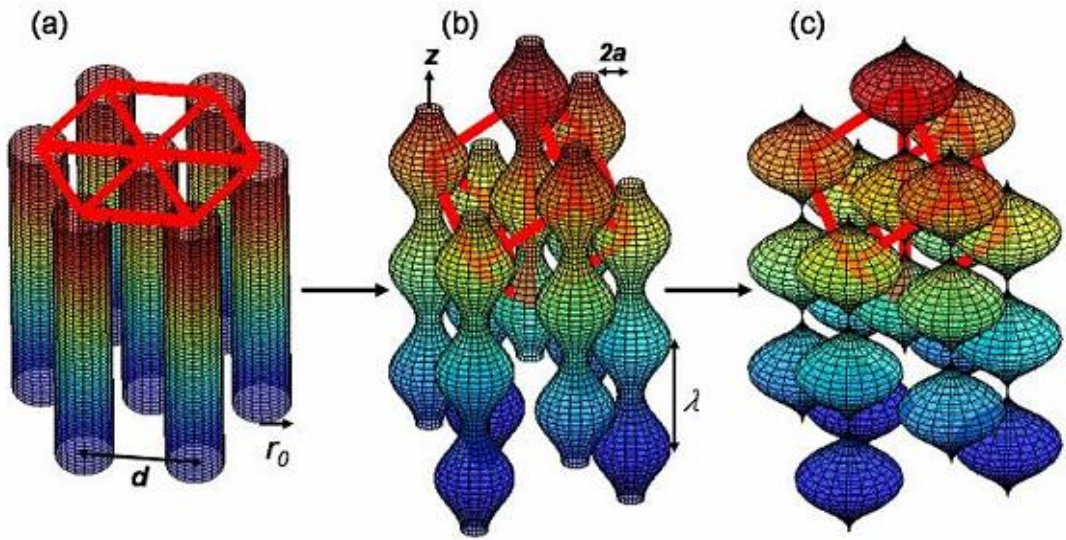

Figure 14. Schematic illustration of the transition from HEX to BCC. (a) The initial state of seven unmodulated HEX cylinders. (b) At the intermediate stage, the cylinders are modulated in a coupled way as discussed in the text. As the amplitude of the modulation $A$ grows the rippled cylinders break up into spheres as shown in (c). When the distance between neighboring cylinders approaches $d = 2\sqrt{2}\lambda/3$, a commensurate BCC structure is formed. A BCC cube is shown as a visual aid.

To represent the ripple due to the anisotropic fluctuations we modulate the radius of a cylinder oriented along the z direction by a transverse wave along the z axis as,

$$r(z) = r_0 + A \cos(2\pi z / \lambda + \varphi). \quad (2)$$

Here $r$ is the radius of cylinder, $A$ and $\lambda$ are modulation amplitude and wavelength respectively, and $\varphi$ is the phase of the modulation. The epitaxial relation for the HEX to BCC transition requires that the <001> direction corresponding to the axis of the cylinder becomes the <111> direction of the BCC lattice, and the three (100) planes of HEX transform into the (110) planes of BCC structure. In order to obtain the epitaxial relationship the modulation of cylinders has to be coupled as shown in Figure 14b.

The question arises as to how to select the phase shifts $\varphi$ between neighboring cylinders. Intuitively it is obvious that if the bulges of two neighboring cylinder are in phase then there will be unfavorable steric interactions. We formalize this idea by a simple calculation of minimizing the overlap volume between neighboring modulated cylinders. Obviously the configuration with minimum overlap volume is the most favorable. The overlap volume shown in Figure 15 was calculated by considering three adjacent cylinders on an equilateral triangular lattice. If we set one of the 3 cylinders as having phase shift 0, then the other two have phase shifts $\varphi$ and $2\varphi$ respectively. This implies a constant difference between neighboring cylinders counted in a cyclic manner. The total overlap volume for the system is $N$ times of that obtained for this triangular unit, where $N$ is the total number of HEX cylinders. To calculate this volume we assume that the cylinders can be sectioned as hard-discs with a radius varying in z direction as given by Eq. (2). In the block copolymer SEBS in mineral oil, the hard-disk radius $r_{hs}$ consists of two parts: $r_0$ which corresponds to the radius of the PS cylindrical core, and $r_{hs} - r_0$ which represents the swollen corona formed by the loops and bridges of the PEB chains. In this sense the hard disk represents the exclude volume interaction. By integrating over the length of the cylinder we obtain the overlap volume as a function of $\varphi$.

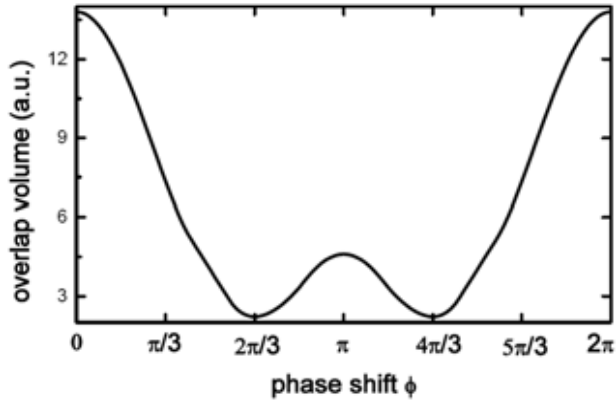

Figure 15. Overlap volume of 3 adjacent modulated cylinders in an equilateral triangle lattice with phase shifts 0, $\varphi$, and $2\varphi$ respectively. The model parameters are set as: $r_{hs}$ = 15.5 nm, $\lambda$ = 33 nm, $A$ = 6.5 nm, $d$ = 31.1 nm and length of cylinder $L$ = 1000 nm. The minima occur at $\varphi = 2\pi/3$ and $4\pi/3$, which are equivalent to each other.

The results obtained by numerical integration, shown in Figure 15, clearly indicate minima at $\varphi$ = $2\pi/3$ and $4\pi/3$. We note that these two values are equivalent[45]. If we alternate the phase shift $\varphi$ of the 6 surrounding cylinders as $4\pi/3$ and $8\pi/3$ with respect to the center cylinder ($\varphi = 0$), then the resulting spheres will arrange on a BCC lattice (Figure 14b), and the epitaxy is automatically satisfied. As the amplitude of the ripple grows and reaches the maximum, $A = r_0$, modulated cylinders break into spheroidal "pearls" as shown in Figure 14c. When the spacing between neighboring cylinders, $d = 2\sqrt{2}\lambda/3$, a commensurate BCC structure is formed with lattice constant $= 2\lambda/\sqrt{3} = \sqrt{1.5}d$. To keep the volume of rippled cylinder conserved as the amplitude of the modulation, $A$, grows, $r_0$ decreases as

$$r_0(A) = \sqrt{r_0^2(A=0) - A^2[1 - \sin(2\pi L/\lambda)/(2\pi L/\lambda)]/2}, \qquad (3)$$

where $L$ is the length of rippled cylinder. We note that an FCC phase will form if we use $d = \sqrt{2}\lambda/6$ and the same phase shifts as above. The yield of the twinned BCC transformed from HEX has been predicted theoretically[46] and demonstrated experimentally[6,7]. According to our model, the twinned BCC arises from two sets of rippled cylinders with clockwise and counter-clockwise phase shifts of ($0$, $4\pi/3$, $8\pi/3$) which are mirror to each other along a HEX (100) or (110) plane.

**Scattering Intensity Calculation.** The calculation of the scattering function of HEX modulated cylinders proceeds in three steps. First we calculate the form factor $p(\vec{q})$ of a single rippled cylinder with a definite orientation relative to the scattering wavevector $\vec{q} = \frac{4\pi}{\lambda}\sin(\frac{\theta}{2})$ where $\theta$ is the scattering angle. Next we calculate the scattering intensity from an oriented domain using the form factor of rippled cylinder and the structure factor of the HEX lattice, and finally obtain the azimuthally averaged scattering intensity by averaging over all orientations.

The form factor of a single rippled cylinder oriented along the z-axis using cylindrical coordinates, is obtained by integrating over the volume $V_{cyl}$ of the rippled cylinder

$$p(\vec{q}) = \int_{V_{cyl}} \exp(-i\vec{q}\cdot\vec{r})\, d^3\vec{r}. \qquad (4)$$

The integration in Eq. (4) over the circular polar coordinates can be performed analytically, giving

$$p(\vec{q}) = \int_{-L/2}^{L/2} \exp(-iqz\cos\alpha) \frac{2\pi(r_0 + A\sin(2\pi z/\lambda))}{q\sin\alpha} B_1[q(r_0 + A\sin(2\pi z/\lambda))\sin\alpha]\, dz, \qquad (5)$$

where $B_1(x)$ is the first order Bessel function, $\alpha$ is the angle between $\vec{q}$ and the z-axis (the polar angle in cylindrical coordinates), and $q$ is the magnitude of $\vec{q}$.

The structure factor $S(\vec{q})$ of a HEX lattice of $N$ cylinders all oriented at an angle $\alpha$ relative to $\vec{q}$ is given by:

$$S(\vec{q}) = 1 + \sum_{i=1}^{N-1} \exp(-i\vec{q} \cdot \Delta \vec{r_i}), \quad (6)$$

where $\Delta \vec{r_i} = \vec{r_i} - \vec{r_0}$ denotes the position vector of the $i$-th cylinder $\vec{r_i}$ relative to a chosen cylinder denoted as $\vec{r_0}$ in the HEX array. For one crystalline domain all the cylinders make the same angle $\alpha$ with $\vec{q}$ so the contribution of a single domain with $N$ cylinders to the scattering is $|S(\vec{q}) \cdot p(\vec{q})|^2$. Since the experimental data described earlier is for unoriented samples and we observed a uniform azimuthal distribution of the scattered intensity we assume that the crystalline domains are randomly oriented. Hence, the azimuthally averaged scattered intensity $I(q)$ is calculated by numerical integration over the angular space as:

$$I(q) = \frac{1}{2} \int_0^\pi |S(\vec{q}) \cdot p(\vec{q})|^2 \sin\alpha \cdot d\alpha. \quad (7)$$

The parameters for the size and spacing were obtained from the experimental data. The radius of the cylinder $r_0$ was taken as 8 nm, and that of the sphere as 10 nm. The spacing between neighboring cylinders $d$ was determined from the peak position using Eq. (8). Since it is not possible to determine the length of the cylinder from the SAXS data, we chose $L = 1000$ nm. This value is of the same order as usually seen in TEM images of block copolymers. The number of cylinders $N$ in one domain was chosen to get the width of $I(q)$ in reasonable agreement with experiment. The larger the number of cylinders in one domain the narrower is the calculated peak. We found reasonable agreement for widths with $N = 381$. A typical calculation based on the model described above is shown in Figure 16 with $A = 4.5$ nm, $\lambda = 32.9$ nm, $d = 32.3$ nm.

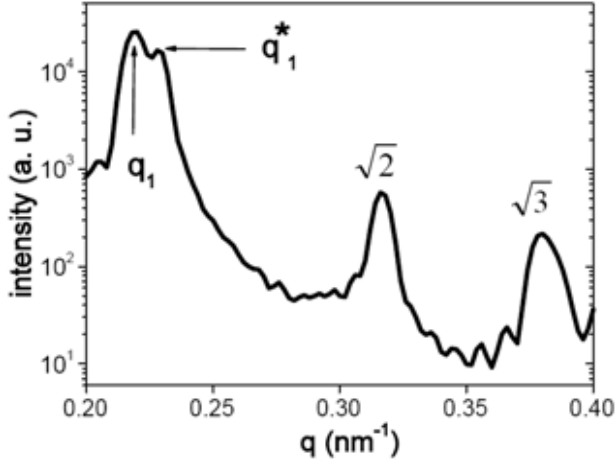

Figure 16. A typical scattering intensity calculated from the model. The parameters of the calculation are $N = 381$, $L = 1000$nm, $r_0 = 8$nm, $\lambda = 32.9$nm, $A = 4.5$nm and $d = 32.3$nm. The main peak ($q_1$) and the side-peak (indicated as $q_1^*$) both contribute to the scattering intensity of the primary peak. The $\sqrt{2}$ peak and $\sqrt{3}$ peak are also indicated.

The results are qualitatively unchanged on varying $L$, $\lambda$, $d$ and $N$, although the peak positions and intensities changed with $d$, $A$ and $\lambda$. The primary peak, the $\sqrt{2}$ peak and $\sqrt{3}$ peak are clearly displayed in Figure 16.

We note that the primary peak is split into two peaks, a main peak ($q_1$) and a side-peak ($q_1^*$). The main peak arises from HEX (100) plane with peak position

$$q_1 = 4\pi/(\sqrt{3}d). \quad (8)$$

The three HEX (100) planes will become the three BCC (110) planes *parallel to the cylinder axis*. The side peak arises from the other three BCC (110) planes that are *not parallel* to cylinder axis with peak position

$$q_1^* = 2\pi\sqrt{\frac{4}{9d^2} + \frac{1}{\lambda^2}}. \quad (9)$$

According to our model, the three BCC (100) planes *parallel to the cylinder axis* will not be identical to the three *non-parallel* BCC (110) planes unless the condition $d = 2\sqrt{2}\lambda/3$ is satisfied and modulated cylinders fully break up into BCC spheres. In other words, the main and side peak will not coincide until $d$ equals $2\sqrt{2}\lambda/3$. Furthermore, the intensity of the side peak $q_1^*$ is weaker than that of main peak $q_1$ and it grows as the modulation amplitude $A$ increases. Eventually when the amplitude $A$ approaches the maximum value $r_0$, i.e., when the modulated cylinders fully break up into spheres, the intensity of the side peak equals that of the main peak. This phenomenon of two principal peaks due to the mismatch of $d$ and $\lambda$ has been addressed by Matsen[23] and can be observed in the experimental data on SIS melt reported by Ryu and Lodge[7]. The side peak is clearly seen in Fig. 7 of reference 7. From the positions of the two peaks in that figure, we can calculate $d$ and $\lambda$ for the SIS melt using Eqs. (8) and (9). We obtain $\lambda \approx 35 nm$ which exactly agrees with their TEM observation. In fact, the side peak $q_1^*$ corresponds to the two sets of 6 fluctuation spots (total 12 spots for the twinned BCC) in a reciprocal space sphere developed by Qi and Wang[46], and the main peak $q_1$ corresponds to 6 spots of original HEX principal peak. However, the BCC structure would produce the same scattering pattern but all spots would be equally intense because they come from identical reflections. In contrast, for modulated cylinders the peak position is mismatched and the intensity of fluctuation spots is weaker than that of the original HEX principal peak. Fig. 14 of reference 7 also shows the appearance of 4 new weak spots in the x-z plane in q space, which grow in intensity with annealing time, providing a clear signature of the formation and growth of modulated cylinders.

The behavior of primary peak width of SAXS data with T-jump $\Delta T = 45$ °C shown in Fig.12b is an indication of the 2-peak splitting. The emergence of the second peak $q_1^*$ (related to $d$ and $\lambda$) broadens the primary peak because the position of the second peak does not coincide exactly with the first peak due to the mismatch of $d$ and $\lambda$. As the modulation amplitude $A$ increases, the second peak grows, meanwhile, the mismatch of $d$ and $\lambda$ decreases as $d$ approaches to $\sqrt{2}\lambda/1.5$. Thus, as $A$ increases, the primary peak first becomes broader and then narrows later, which is consistent with the experimental result shown in Fig.12b.

Due to the resolution limit of our experiment, we were not able to resolve the two-peak splitting visually from the SAXS data. However, the peak profile was asymmetric as shown in Figure 17a. To confirm that this asymmetry is due to a second peak close to the primary peak, we used a simple procedure of reflecting the data below the maximum position and then subtracting the reflected curve from the original data. As seen from Figure 17a, the subtracted intensity indicates a second peak whose intensity grows with time. Due to the uncertainty of the choice of the reflection position, this method is not suitable to determine the position of the two peaks quantitatively. To obtain the peak positions we fit the primary peak of the experimental data with 2 Gaussian peaks (shown in Figure 17b). The time-evolution of $d$ and $\lambda$ (shown in Figure 17c) was determined from the positions of the two peaks using Eqs. (8) and (9). Figure 17b shows that the two initially indistinguishable peaks (within the experimental resolution) separate with increasing time and then eventually merge together. Note that at $t = t_2 = 200$ s, when BCC is formed, the two peaks are not merged indicating that $d$ and $\lambda$ do not satisfy the commensuration relationship at this time. The spheres continue to move eventually forming a well-defined BCC structure with $d$ = 31 nm at around 1500 s, which agrees with the prediction by Matsen[23]. We note that in calculating $d$ and $\lambda$ the peak with higher $q$ value is identified as the side peak $q_1$* because the corresponding intensity is lower than of the other one and grows as time increases.

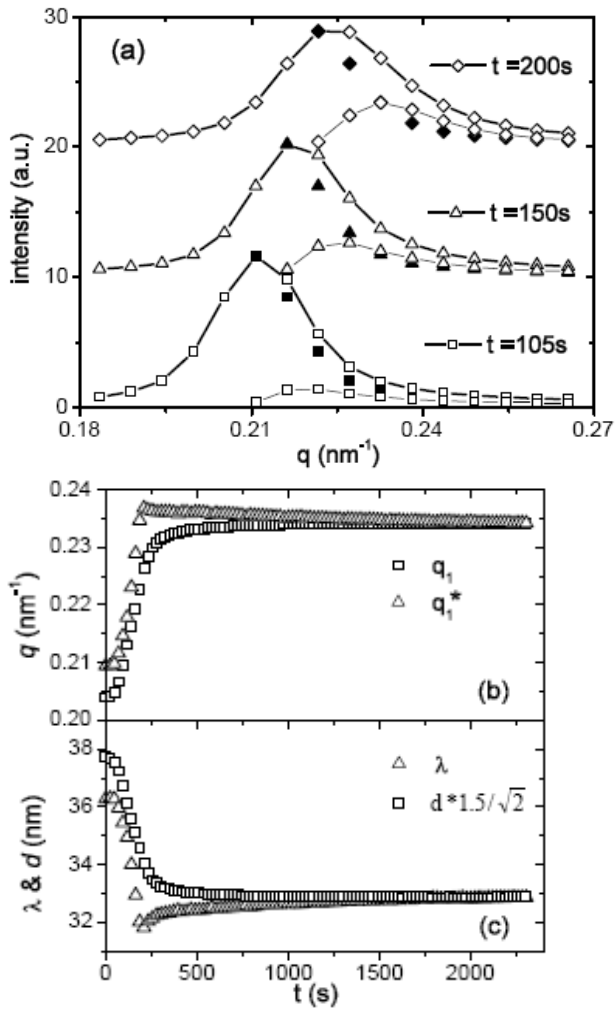

Figure 17. (a) The primary peak profile of a few frames of SAXS data for T-jump $\Delta T = 45\ ^oC$ (empty symbols and thick lines). Also shown are the mirror reflections (filled symbols) and the subtractions (empty symbols and thin lines). Frames at different times are shifted for clarity. (b) The time evolution of the positions of the two peaks obtained by fitting the primary peak with 2 Gaussians. (c) $\lambda$ and $d$ obtained from the peak positions using Eqs. (8) and (9). To illustrate the effect of commensuration $d$ is multiplied by $1.5/\sqrt{2}$.

**Comparison of Observed Kinetics with the Model Calculation.** As we addressed before, our model is best suited to explaining the temperature jump beyond the spinodal ($\Delta T$ = 45 °C), where all the cylinders ripple simultaneously. The situation is more complicated for the nucleation and growth scenario (shallow temperature jump below the spinodal) because some parts of cylinders would develop ripples while others would remain unmodulated as discussed by Matsen[23] and the front would advance with time.

In order to compare the model to the experiment with T-jump $\Delta T$ = 45 °C we use the values of $d$ and $\lambda$ obtained from the two Gaussian peaks fitting procedure (see Figure 17 and Table 3).

Table 3. The parameters of the rippled cylinders ($A$, $d$, and $\lambda$) used for model scattering intensity calculation.

| $A$ (nm) | 0 | 1 | 2.5 | 3.5 | 4 | 4.5 | 5 | 5.8 | 6.5 | 6.5 |
|---|---|---|---|---|---|---|---|---|---|---|
| $d$ (nm) | 35.6 | 34.3 | 33.8 | 33.6 | 33.3 | 33.1 | 32.8 | 32.6 | 32.3 | 31 |
| $\lambda$ (nm) | -- | 35.2 | 34.5 | 34 | 33.5 | 32.9 | 32.4 | 32 | 31.8 | 32.8 |

There is no direct way to obtain the amplitude as a function of time from the data. As a simple approach the value of $A$ is set such that during the transition period ($t_0 < t < t_1 = t_2$), $A$ increases roughly linearly from 1 nm to 6.5 nm ($r_0$ decreases to 6.5 nm as $A$ = 6.5 nm). The results of the numerical calculation of the scattering intensity are shown in Figure 18a along with SAXS data for the T-jump $\Delta T$ = 45°C.

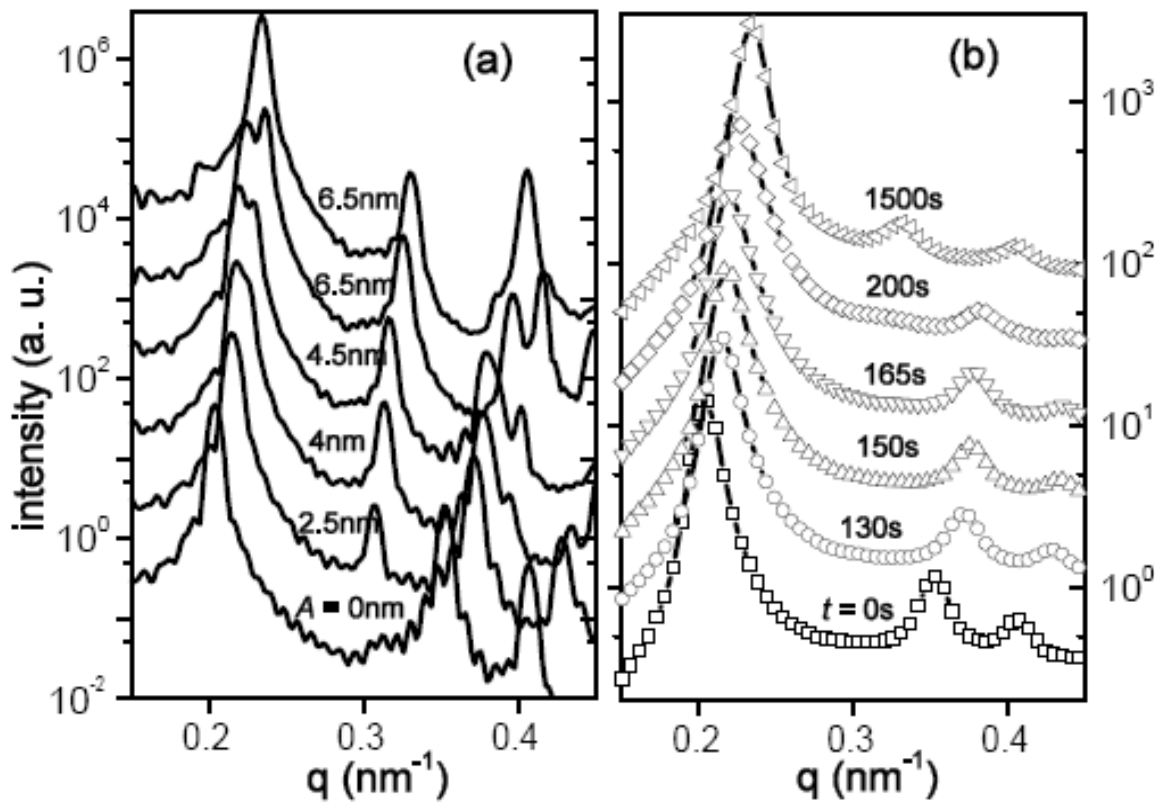

Figure 18. (a) The calculated scattering intensity at different values of the modulation amplitude $A$ as indicated. The parameters of calculation are $N$ = 381, $r_0$ = 8 nm, $L$ = 1000 nm. The values of $\lambda$ and $d$ vary as $A$ varies from 0 to 6.5 nm (see Table 3). (b) Selected frames from $t$ = 0 to 1500 s as indicated, from the time-dependent SAXS data for T-jump $\Delta T$ = 45°C are shown for comparison with the calculation. The scattering curves are shifted vertically for clarity.

The splitting of the primary peak of the calculation is clearly seen for the calculations with $A$ = 4 - 6.5 nm. The positions of the two peaks $q_1$ and $q_1^*$, as well as their integrated intensity (defined as the product of peak intensity and width) $I_1$ and $I_1^*$ obtained by a Gaussian fitting procedure are plotted in Figure 19. The integrated intensity of the side peak $I_1^*$ increases with increasing $A$, whereas $I_1$ decreases. When $A = r_0 = 6.5$ nm, the intensities are equal to each other. The dependence of the intensity on the amplitude agrees with the self-consistent field calculation of Matsen[23] (see Fig. 4 and 7 of ref. 23).

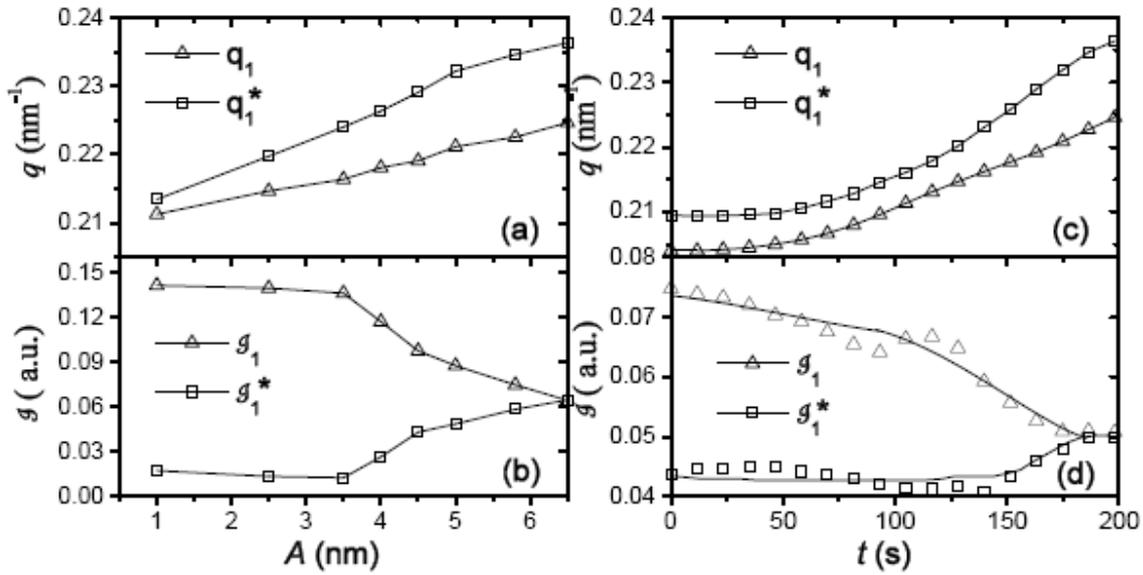

Figure 19. (a) Peak positions $q_1$ and $q_1^*$, and (b) integrated intensities $I_1$ and $I_1^*$ obtained from 2-Gaussian fit of the primary peak for the model calculation (Figure 18a) as a function of the modulation amplitude $A$. For comparison $q_1$, $q_1^*$, and $I_1$, $I_1^*$ obtained from the SAXS data for the T-jump with $\Delta T = 45$ °C at early times are shown in (c) and (d). The solid lines are shown for guide to eye.

Overall, the numerical calculation of scattering intensity of the model agrees well with the SAXS data for T-jump $\Delta T = 45$°C (shown in Figure 18b). We have not attempted a fit of the data to the model calculation because of the uncertainties of determining $L$ and the time dependence of the amplitude. From the calculation (Figure 18a), we can observe a very clear appearance of √2 peak even at small amplitude $A$. Experimentally, the onset of √2 peak could signal the formation of modulated cylinders with ripples at BCC symmetry and does not necessarily indicate spheres in BCC phase.

**Conclusions**

We have examined the kinetics of HEX to BCC transition in the triblock copolymer SEBS 45% in mineral oil, a selective solvent for the middle PEB block, using time-resolved SAXS measurements. Temperature-ramp SAXS data show that the HEX to BCC transition occurs at ~127 °C with the spinodal $T_s$ at ~150 °C, and ODT at ~180 °C. By examining various T-jumps with the sample initially at 110 °C, we were able to observe the nucleation and growth mechanism driven kinetics for shallow T-jump and the spinodal decomposition with continuous ordering for deep T-jump. Temperature-jump experiments starting from 110 °C show that the nucleation and growth kinetics involves three stages. In the first stage, $t_0 < t < t_1$, the cylinders get close to each other while remaining in HEX structure; the second stage, $t_1 < t < t_2$, is the transition period, where the cylinders are modulated along their axis and eventually break into spheres on a BCC lattice, and finally in the third stage, $t > t_2$, the domains coalesce and the fraction of material in the BCC state grows.. The transition time, $t_2 - t_1$, decreases linearly with increasing $\Delta T$ and extrapolates to zero at $\Delta T = 40$ °C, corresponding to a spinodal at 150 °C for the metastability limit of HEX in BCC. For a deep T-jump to a temperature above the spinodal, the first two stages merge together, and the HEX to BCC transition occurs via a mechanism involving continuous ordering and spinodal decomposition. In this case, after the initial temperature equilibration time $t_0$, the HEX cylinders transform to BCC spheres until $t = t_2 (= t_1)$ and after that the BCC domains coalesce and grow. We also examined the kinetics of the HEX to disordered spheres transition and observed that the system first transforms from HEX to BCC, followed by the order-disorder transition.

To calculate the scattering during the transformation stage we have developed a geometrical model based on the previous theoretical models of anisotropic fluctuations, according to which the cylinders develop a transverse wave-like instability that grows with time leading to the formation of spheres. We found that when the phase shift $\phi$ of 3 adjacent cylinders in the unit cell are (0, $4\pi/3$, $8\pi/3$) the overlap volume is minimized and the centers of the spheroidal bulges lie on a BCC lattice. This model automatically preserves the epitaxial relationship observed in experiments, i.e., the cylinder axis becomes the <111> direction of BCC, and the (100) planes of HEX transform into the (110) planes of BCC. The calculated scattering intensity of the model agrees well with the time-resolved SAXS data for the T-jump above the spinodal. We found that initially the wavelength $\lambda$ of the modulation is incommensurate with the cylinder spacing $d$. This leads to a splitting of the primary peak into two peaks which merge together when $\lambda = d*1.5/\sqrt{2}$. The integrated intensity of the higher-$q$ component increases while that of the other one decreases as the modulation amplitude increases; becoming equal to each other as the modulation approaches its maximum. Although the two peaks could not be directly resolved in the SAXS data reported here, their presence was inferred from the asymmetric shape of the primary peak. The calculations reported here further support the theoretical predictions from previous studies concerning the mechanism of the HEX to BCC transition.

**Acknowledgements**

This research was supported by NSF Division of Materials Research Grant No. 0405628 to R.B. The SAXS measurements were carried out at Beamlines X27C and X10A of NSLS, Brookhaven National Laboratory which is supported by the U.S. Department of Energy, Division of Materials Sciences and Division of Chemical Sciences, under Contract No. DE-AC02-98CH10886. We thank Dr. Igor Sics and Dr. Lixia Rong of beamline X27C, and Steve Bennett of beamline X10A for technical support at NSLS, and Randy Ewoldt of MIT for help with the rheology measurements. We acknowledge the support of Boston University's Scientific Computation and Visualization group which is supported by NSF for computational resources. We thank Professors Bill Klein and Karl Ludwig for many stimulating discussions.

---

[44] The value of the PS segment length $a_{PS}$ was obtained from Brandrup, J.; Immergut, E.H., Eds.; *Polymer Handbook*, 2nd ed. Wiley-Interscience: New York, 1975; p IV-40.

[45] Phase shifts of 3 cylinders (0, *2π/3*, *4π/3*) and (0, *4π/3*, *8π/3*) are equivalent because *8π/3=2π/3+2π*.

[46] Qi, S.; Wang, Z.-G. *Polymer* **1998,** *39*, 4639.